\documentclass[aps,prb,twocolumn, superscriptaddress, amsmath, amssymb]{revtex4-2}

\usepackage{xcolor}
\usepackage[normalem]{ulem} 
\usepackage{cancel}

\usepackage{graphicx}  

\usepackage[utf8]{inputenc}  

\usepackage[english]{babel} 
                                   
\usepackage{dcolumn}	
\usepackage{bm}  



\begin{document}

\title{Massless Dirac Fermions in curved surfaces with localized curvature}

\author{A. R. N. Lima}
\email{romariolimafisica@alu.ufc.br}
\affiliation{Universidade Federal do Ceará, Departamento de Física, Fortaleza, CE, Brazil}

\author{D. F. S. Veras}
\email{diego.veras@ufca.edu.br}
\affiliation{Universidade Federal do Cariri, Centro de Ciências e Tecnologia, Juazeiro do Norte, CE, Brazil}

\author{J. E. G. Silva}
\email{euclides@fisica.ufca.br}
\affiliation{Universidade Federal do Ceará, Departamento de Física, Fortaleza, CE, Brazil}

\begin{abstract}
We investigate how a localized curvature affects the dynamics of massless Dirac fermions in a curved surface. We consider a smooth bump with axial symmetry, adopting two specific geometric models, namely a Gaussian  and a volcano-like bumps. By considering a minimal coupling between the spinor and the surface geometry, described by the vielbeins and the spin connection, we study the behavior of the wave function over the surface.  By using appropriate numerical methods, we find a linear discrete energy spectrum for the Dirac fermions and its corresponding wavefunctions when the Fermi velocity is considered. It turns out that, since the curvature vanishes asymptotically, the electron states are free waves far from the bumps, but around the curved points, the wave function increases its probability density.

\end{abstract}

\maketitle

\section{Introduction}
\label{sec:intro}

In recent years, two-dimensional systems such as graphene, a single layer of carbon, have attracted significant interest due to its exceptional electronic, mechanical, and thermal properties \cite{geim2007graphene,geim2009graphene}. One of these properties is the emergence of smooth ripples due to the interplay between internal stresses and structural fluctuations \cite{seung1988defects}. These ripples, which can reach heights of a few angstroms and lengths of several nanometers \cite{de2007charge,sutter2009scanning}, have been observed in suspended graphene samples through experimental techniques such as Transmission Electron Microscopy (TEM) \cite{meyer2007roughness,meyer2007structure} and Scanning Tunneling Microscopy (STM) \cite{sutter2009scanning,ishigami2007atomic}. These geometric deformations in graphene modify the effective Hamiltonian, leading to position-dependent Fermi velocities and generating pseudogauge fields \cite{monteiro2023dirac,de2012space,vozmediano2008gauge}.

The curvature of the surface directly influences the dynamics of the fermions; for example, smoothly curved regions locally modify the components of the Pauli matrices, inducing a local modulation of the Fermi velocity \(v_F\), which ceases to be constant and becomes position‑dependent, generating fluctuations in the density of states \cite{de2007charge}.Local anisotropies and charge inhomogeneities, the formation of effective wells or barriers favoring the emergence of bound states, and modifications of the effective potential are examples of curvature-induced effects on the dynamics of fermions. \cite{hayashi2010curvature,batista2018curvature,kothari2019critical,yan2013strain,yang2012electronic}. These and other properties have also been explored in geometries such as the cone \cite{furtado2008geometric}, Möbius strip \cite{monteiro2023dirac}, helicoidal surface \cite{atanasov2015helicoidal,sutter2009scanning}, catenoid bridge \cite{silva2020electronic, yecsiltacs2022dirac}, and the Gaussian surface \cite{de2007charge}.  

Inspired by these works, we investigate how a localized curvature modifies the dynamics of electrons intrinsically defined on curved surfaces. More specifically, we explore how these bumps affect the observable electronic states, both in terms of the energy spectrum and the local probability density. Initially, we couple the Dirac equation to the curved geometry by adopting a minimal coupling with the surface vielbein and spin connection, so as to consistently incorporate the effects of curvature into the kinetic operator. We employ two axially symmetric surfaces: a Gaussian bump and a volcano‑type surface, whose radial profiles differ by the presence of a central valley or a maximal apex at \(r=0\). Axial symmetry implies conservation of total angular momentum in the $z$-direction, and thus the Dirac equation can be reduced to a one‑dimensional problem along the radial direction, which simplifies both the analytical treatment and the numerical computation. We find that both bump geometries produce non‑localized states — i.e., states spread out in space — whose probability densities show significant accumulations in the bump regions, suggesting that curvature acts as an effective potential capable of attracting or repelling fermions depending on the surface topography.

In addition, we introduce an external magnetic field to analyze the combined effects of the pseudogauge field generated by curvature with a real magnetic field. We observe that, in the absence  of an external field, there are no bound states, whereas the introduction of the magnetic field leads to bound Landau levels \cite{bueno2012landau}. This occurs because the magnetic field alters the effective potential asymptotically, leading to energy quantization \cite{villalba2001energy}. Thus, this work also aims to investigate how this external field modifies the probability densities on the surfaces.

The present work is organized as follows: In Section II, we discuss the massless Dirac equation in (2+1) dimensions on a curved surface with axial symmetry; we define the metric, explicitly calculate the Christoffel symbols, introduce the Vierbein field, and construct the Dirac Hamiltonian for geometric configurations with this symmetry; at the end of this section, we present the two geometries of interest, the Gaussian surface and the volcano-like surface. In Section III, we analyze the dynamics of fermions in these curved geometries, comparing the Gaussian and volcano surfaces through effective potentials and probability densities of finding the electron in each graphene sublattice.  In Section \ref{Sec-Numerical} we perform a numerical analysis of the fermion energy spectrum and the corresponding eigenfunctions for both surfaces considered in the paper.   Then, in Section IV, we include a constant external magnetic field and demonstrate how bound states and Landau levels emerge in this configuration. Finally, in Section V, we present our final considerations, highlighting the main conclusions and perspectives of this work.


\section{Massless Dirac equation in (2+1) dimensions on a curved surface with axial symmetry.}
\label{sec2}

To understand the influence of geometry on the dynamics of electrons, it is essential to construct the Hamiltonian using the covariant formalism, which incorporates the metric induced by the embedded surface. We first derive the Dirac Hamiltonian for flat space, subsequently, we will analyze how this Hamiltonian is modified by the presence of curvature. For this purpose, it is essential to determine a general metric that describes a smooth and properly coupled bump, free of singularities on the graphene sheet.

The surface is described by a function  \(z(r)\), which is asymptotically flat in the $(x,y)$ plane and describes the height relative to the flat surface (\(z = 0\)) \cite{de2007charge}. The spacetime metric is written as
\begin{equation}
	ds^2= dt^2-dr^2-r^2d\theta^2-dz^2.
\end{equation}
 Using $z=z(r)$, we have
\begin{equation}
	dz^2 = \left( \frac{dz}{dr} \right)^2 dr^2 = \alpha f(r) dr^2,
\end{equation}
where \(f(r)\) characterizes the radial dependence of the curvature. Consequently,
\begin{equation}
	ds^2=dt^2-(1+\alpha f(r))dr^2 -r^2d\theta^2,
\end{equation}
where $\alpha$ is a dimensionless perturbation parameter defined by
\begin{equation}
	\alpha= \frac{A^2}{b^2}.
\end{equation}	
 The resulting spatial metric, including the curvature term,, is expressed as
\begin{equation}
	g_{ij} = 
	\begin{pmatrix}
		-(1 + \alpha f(r)) & 0 \\
		0 & -r^2
	\end{pmatrix}. \label{Eq.5}
\end{equation}
 In the absence of curvature ($\alpha=0$), the spatial metric reduces to the polar-coordinate metric: 
\begin{equation}
	g_{ij} =
	\begin{pmatrix}
		-1 & 0 \\
		0 & -r^2
	\end{pmatrix}.
\end{equation}

To describe the states of electrons (fermions) on a curved surface in graphene, the massless Dirac equation is used:
\begin{equation}
	i\hbar v_F \gamma^{\mu} (\partial_{\mu} + \Omega_\mu) \psi = 0 \, .
\end{equation}
In this equation, $v_F$ is the Fermi velocity and $ \Omega_\mu$ is the spin connection that ensures the correct definition of the covariant derivative \cite{abergel2010properties,olpak2012dirac,gallerati2019graphene}. The matrices $\gamma^{\mu}$ incorporate the relativistic structure of the fermions and  they must satisfy the anticommutation relation (or Clifford algebra):
\begin{equation}
	\{ \gamma^{\mu}, \gamma^{\nu} \} = 2 g^{\mu \nu}.
\end{equation}

The spin connection $\Omega_\mu$ is constructed from the vierbeins and is expressed as
\begin{equation}
	\Omega_\mu=  \frac{1}{8} \omega_{\mu}^{ab} [\gamma_a,\gamma_b] \, ,
\end{equation}
where
\begin{equation}
	\omega_{\mu}^{ab} = e_{\nu}^a (\partial_{\mu} e^{\nu b} + \Gamma_{\mu \lambda}^{\nu} e^{\lambda b}) \, .
\end{equation}
Here, the affine connection (Christoffel symbols)  $\Gamma_{\mu \nu}^{\rho}$ are related to the variation of the metric and are defined by
\begin{equation} 
	\Gamma_{\mu \nu}^{\rho} = \frac{1}{2} g^{\rho \lambda} \left( \partial_{\mu} g_{\nu \lambda} + \partial_{\nu} g_{\mu \lambda} - \partial_{\lambda} g_{\mu \nu} \right).
\end{equation}
When working with polar coordinates in the flat case (no curvature), the non-zero Christoffel symbols are
\begin{equation}
	\Gamma_{\theta\theta}^r = -r  \qquad  \text{and} \qquad \Gamma_{r\theta}^\theta = \Gamma_{\theta r}^\theta = \frac{1}{r} \, .
\end{equation}

The vierbein fields \( e^a_{\mu} \) establish the relationship between the curved metric  \( g_{\mu\nu} \) and the flat metric  \( \eta_{\mu\nu}\) through the relation \cite{birrell1984quantum}: 
\begin{equation}
g_{\mu\nu} = e^a_{\mu} e^b_{\nu} \eta_{ab} \, , 
\end{equation}
where \( \eta_{ab} \) is given by
\begin{equation}
	\eta_{ab} = 
	\begin{pmatrix}
		-1 & 0 \\
		0 & -1
	\end{pmatrix}.
\end{equation}
This relation does not uniquely determine the vierbeins  \( e^a_{\mu} \), allowing two natural choices that represent local flat frames. The first choice,
\begin{equation}
e^a_{\mu} = (e^\mu_a)^{-1} = 
\begin{pmatrix}
1 & 0 \\
0 & r
\end{pmatrix} \ ,
\end{equation}
 preserves the Cartesian form of the gamma matrices, which results in a constant gauge connection \cite{de2007charge}. On the other hand, the second choice,
\begin{equation}
e^a_{\mu} =
\begin{pmatrix}
\cos\theta & -r\sin\theta \\
\sin\theta & r\cos\theta
\end{pmatrix} \ ,
\end{equation}
corresponds to a frame that rotates with the polar angle $\theta$, modifying the structure of the gamma matrices in such a way that it does not lead to an effective gauge field \cite{shokri2022rkky}. These choices have direct implications for the form of the Hamiltonian obtained later. By adopting the first choice, one of the components of the spin connection is obtained as $\omega_\theta^{12} = -\omega_\theta^{21}=1$, considering the following representation for the Dirac matrices:
\begin{equation}
	\gamma_0 = \sigma_3, \quad \gamma_1 = i\sigma_2, \quad \gamma_2 = -i\sigma_1 \, .
\end{equation}
From this representation and the choice of the vierbeins, the Dirac Hamiltonian in polar coordinates for flat spacetime becomes:
\begin{equation}
H_{\text{flat}} = -i\hbar v_F \begin{pmatrix}
	0 & \partial_r + \frac{i}{r}\partial_\theta + \frac{1}{2r} \\
	\partial_r - \frac{i}{r}\partial_\theta + \frac{1}{2r} & 0
\end{pmatrix}.
\end{equation}

When the curved metric (\ref{Eq.5}) is considered, we obtain the following affine connection coefficients:
\begin{equation}
	\Gamma^r_{rr} = \frac{\alpha f'(r)}{2(1 + \alpha f(r))} \, ,
\end{equation}
\begin{equation}
	\Gamma^r_{\theta\theta} = -\frac{r}{(1 + \alpha f(r))} \, ,
\end{equation}
\begin{equation}
	\Gamma^\theta_{r\theta} = 	\Gamma^\theta_{\theta r} = \frac{1}{r} \, .
\end{equation}
Next, we choose the vierbein matrix in the form
\begin{equation}
	e^{a}_{\mu} =
	\begin{pmatrix}
		(1 + \alpha f(r))^{\frac{1}{2}} \cos\theta & -r\sin\theta \\
		(1 + \alpha f(r))^{\frac{1}{2}} \sin\theta & r\cos\theta
	\end{pmatrix} \, ,
\end{equation}
which incorporates the modification introduced by the curvature through the factor $(1 + \alpha f(r))^{\frac{1}{2}} $. Calculating the spin connection coefficients, we obtain
\begin{equation}
	\omega_\theta^{12} = -\omega_\theta^{21} = 1 - (1 + \alpha f(r))^{-\frac{1}{2}} \, ,
\end{equation}
which implies that
\begin{equation}
\Omega_r = 0 \quad \text{and} \quad\Omega_\theta = \frac{1 - (1 + \alpha f)^{-1/2}}{2} \, \gamma^1 \gamma^2 \, .
\end{equation}
Thus, the Dirac Hamiltonian in curved spacetime can be written as 
\begin{equation}
\hat{H}_C = -i\hbar v_F \begin{pmatrix}
	0 & F(r)\partial_r + \frac{i}{r}\partial_\theta + A_\theta \\
	F(r)\partial_r - \frac{i}{r}\partial_\theta + A_\theta & 0
\end{pmatrix}\, ,
\end{equation}
where we defined
\begin{equation}
    F(r)= (1+\alpha f(r))^{-\frac{1}{2}}
\end{equation}
and
\begin{equation}
	A_\theta= \frac{1- F(r) }{2r} \, . \label{Atheta}
\end{equation}

The term $A_\theta$ can be interpreted as a gauge pseudopotential since its origin is not electromagnetic but rather geometric — it arises due to the curvature of the ripples in the graphene sheet and is intrinsically linked to the spin connection. Thus, this pseudopotential modifies the trajectory of quasiparticles, without representing a real external force, being simply a manifestation of geometric effects associated with the curvature \cite{de2007charge}. 

The curvature scalar
\begin{equation}
R \;=\; g^{\mu\nu}R_{\mu\nu} \, ,
\end{equation}
quantifies the intrinsic curvature of the surface, that is, the local deviation from a flat geometry as ''seen'' by fermions. Starting from the definition of the Ricci tensor \cite{wald2010general}
\begin{equation}
R_{\mu \nu}
=\,
R^\rho{}_{\mu\rho\nu}
=\,
\partial_{\rho}\Gamma^{\rho}_{\nu\mu}
-\partial_{\nu}\Gamma^{\rho}_{\rho\mu}
+\Gamma^{\rho}_{\rho\lambda}\,\Gamma^{\lambda}_{\nu\mu}
-\Gamma^{\rho}_{\nu\lambda}\,\Gamma^{\lambda}_{\rho\mu}\, ,
\end{equation}
we obtain
\begin{equation}
R(r)
\;=\;
-\,\frac{\alpha\,f'(r)}{r\,\bigl[1+\alpha\,f(r)\bigr]^2}\, .
\label{Eq-Escalar_Curvatura}
\end{equation}

Now, in the following subsections, we present two surfaces of interest: the Gaussian surface and the volcano-shaped surface.

\subsection{ The Gaussian surface}
\begin{figure}[h!]
        \centering
        \includegraphics[width=0.9\linewidth]{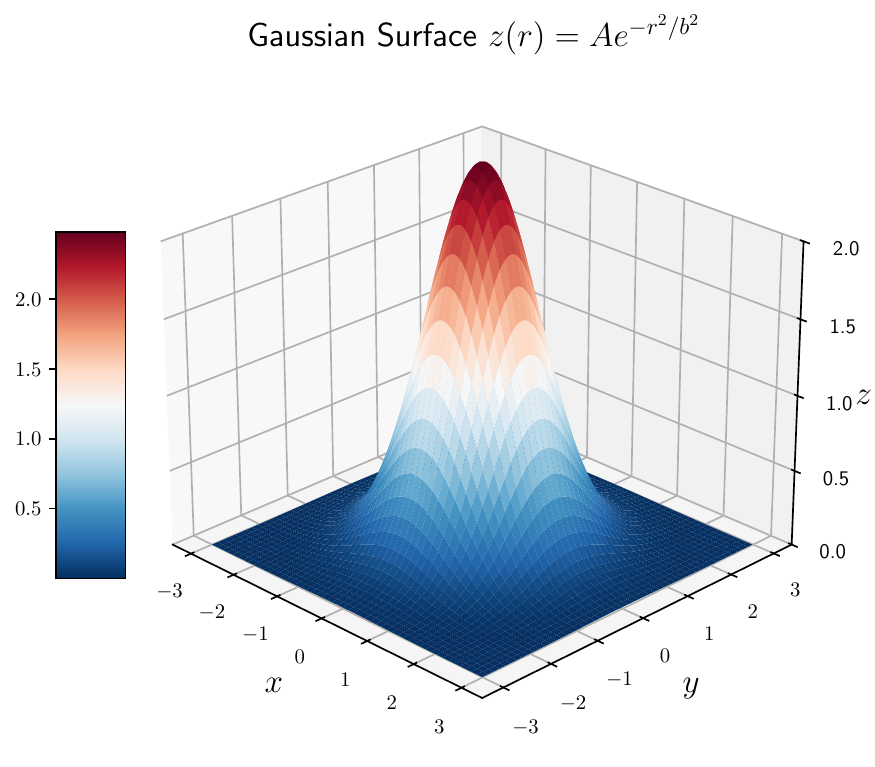}
        \caption{Plot of the Gaussian surface for $A = 2.5$ and $b = 1.25$.}
        \label{Fig-Gaussian_3D}
    \end{figure}

Given that the initial focus is the analysis of a Gaussian surface, we consider the function $z(r)$ defined by  
\begin{equation}
z(r) = A e^{-\frac{r^2}{b^2}},
\end{equation}
where $A$ represents the maximum amplitude of the bump, and $b$ determines its width \cite{de2007charge}. This choice ensures that the elevation is highest at the center of the surface and decreases smoothly as $r$ increases. In the physical context, it is known that $b > A$, as discussed in \cite{de2007charge}, with $b$ typically on the order of a few nanometers and $A$ on the order of a few angstroms. However, since our purpose is to investigate the effects of a localized curvature, we will initially adopt the opposite regime, considering $A > b$, in order to emphasize the geometric effects of the deformation, unless stated otherwise.  We plot in Figure \ref{Fig-Gaussian_3D} the Gaussian surface in a spatial perspective. The parameter $A$  controls the height of the bump while $b$ controls its width.

By calculating the differential term associated with the height variation, we obtain
\begin{equation}
dz^2 = \frac{A^2}{b^4} \, 4r^2 \exp\left(-\frac{2r^2}{b^2}\right) dr^2 \, .
\end{equation}
Thus, the function $f(r)$, which characterizes the radial influence of the curvature on the metric, can be identified as
\begin{equation}
	f_{\text{\tiny G}}(r) = 4 \left( \frac{r}{b} \right)^2 e^{-\frac{2r^2}{b^2}} \, .
	\label{Eq-f(r)_Gaussiana}
\end{equation}

 We plot in Figure \ref{Fig-R(r)_Gaussiana} the scalar curvature for the Gaussian bump and we observe that the scalar curvature exhibits a small negative dip near $r=0$, followed by a positive peak and decays exponentially to zero as $r \to \infty$. In this way, $R(r)$ directly reveals the regions of saddle-like (negative) and dome-like (positive) curvature induced by the graphene topography. { The scalar curvature in Eq. (\ref{Eq-Escalar_Curvatura}), might suggest that it becomes singular for $f(r) = -b^2/A^2$, however, the function $f_{\text{\tiny G}}(r)$ is strictly positive as can be seen from Eq. (\ref{Eq-f(r)_Gaussiana}) and its derivative given by
\begin{equation}
f^{\prime}_{\text{\tiny G}}(r) = \left( \dfrac{2}{r} - \dfrac{4r}{b^2}\right)f(r) 
\end{equation}
is non-singular for all $r>0$.
\begin{figure}[h!]
    \centering
    \includegraphics[width=0.9\linewidth]{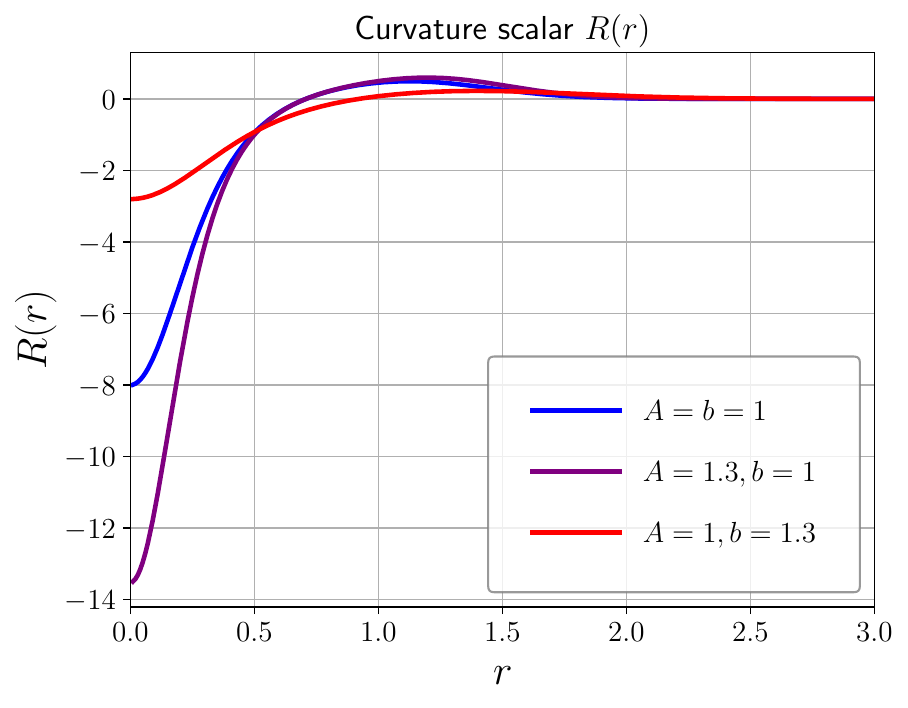}
\caption{Curvature scalar for the Gaussian bump. The plots correspond to the cases $A=b$ (blue), $A>b$ (purple), and $A<b$ (red).}
    \label{Fig-R(r)_Gaussiana}
\end{figure}


\subsection{The volcano-shaped surface}
We now consider another geometry another interesting geometry that also possesses axial symmetry. This configuration corresponds to a ``volcano''-type surface, from which we will determine the function $f(r)$ that characterizes this new geometry. Initially, we define the surface profile by the function
\begin{equation}
z(r) = A r e^{-\frac{r^{2}}{b^{2}}}\, ,
\end{equation}
where $A$ and $b$ are geometric parameters similar to the Gaussian ones. Throughout our analysis, we will work in the regime where $b>A$ which generates the typical ''volcano'' shape, a ring-shaped peak with a central cavity. Although we plot the volcano surface for $b < A$ in Figure \ref{Fig-Vulcao_3D} for a better viewing, more interesting and noteworthy results arise for $b > A$, where the electron ``feels'' the effect of the volcano cavity. For large values of $A$, the numerical results are very similar to those of the Gaussian surface. It is important to note that other regimes do not yield a physically suitable geometry for the problem: when $A \ll b$, the exponential factor decays so rapidly that the surface becomes essentially flat over most of the domain, while for $A \gg b$ a very narrow and sharp peak emerges, which is not representative of a smooth localized curvature. 

\begin{figure}[h!]
    \centering
    \includegraphics[width=0.9\linewidth]{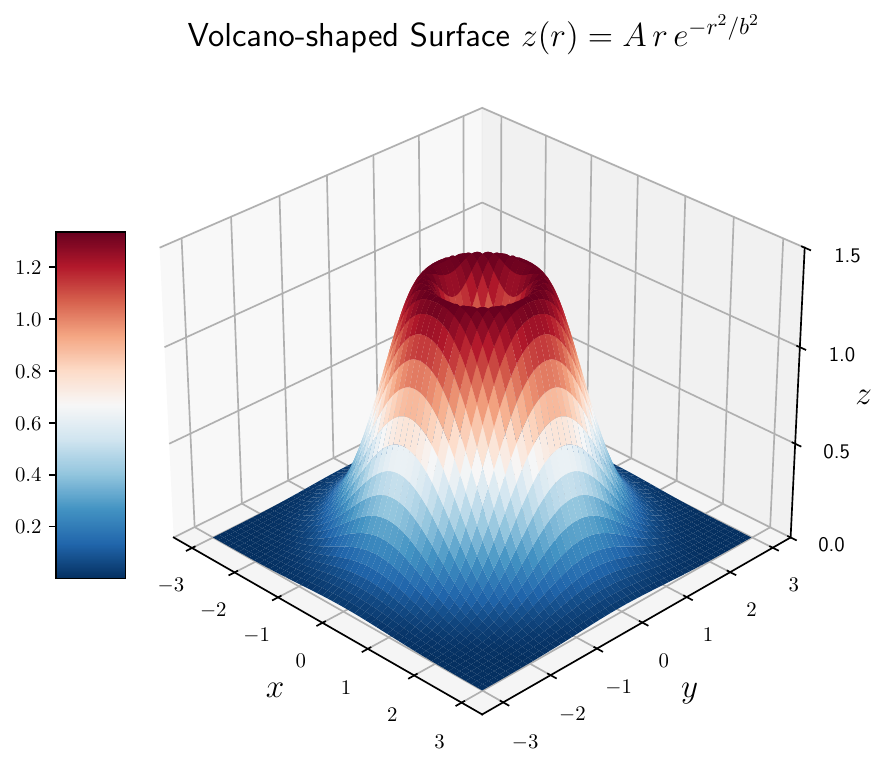}
    \caption{Plot of the volcano-shaped surface for $A = 2.5$ and $b = 1.25$.}
    \label{Fig-Vulcao_3D}
\end{figure}

From this definition, the differential contribution of displacement in the $z$ direction can be expressed as
\begin{equation}
	dz^2 = A^2 e^{-\frac{2r^2}{b^2}} \left(1 - \frac{2r^2}{b^2}\right)^2dr^2 \, .
\end{equation}

Analogously to the Gaussian case, we define the function $f(r)$ such that
\begin{equation}
	f_{\text{\tiny V}}(r) = \left(1 - \frac{2r^2}{b^2}\right)^2 b^2 e^{-\frac{2r^2}{b^2}}.
\end{equation}
This function $f_{\text{\tiny V}}$ captures the geometric effects on the spatial distribution of electrons.

\begin{figure}[h!]
        \includegraphics[width=0.9\linewidth]{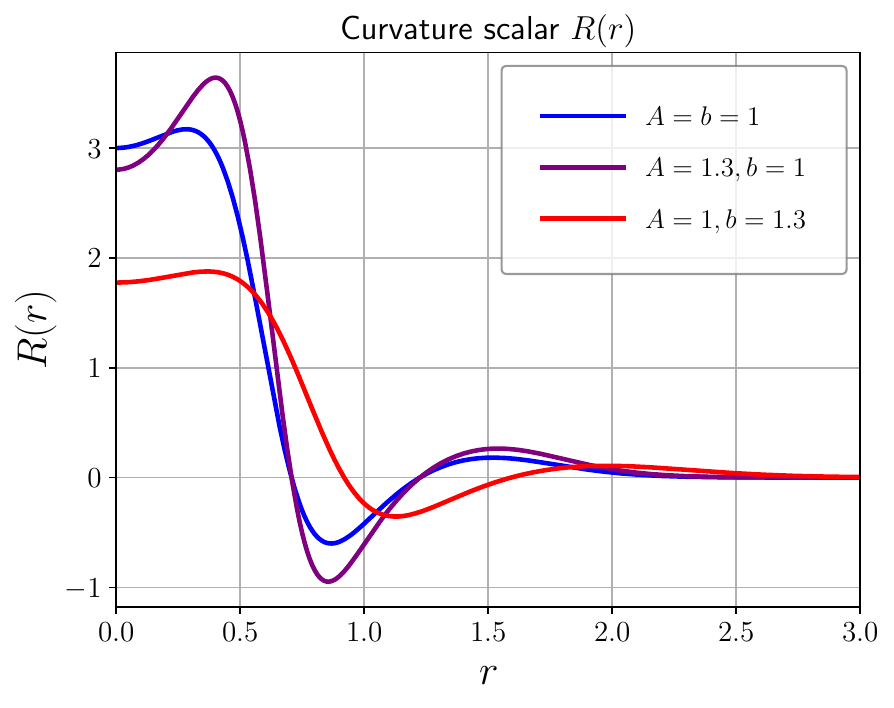}
        \caption{Scalar curvature for volcano-shaped bump. The plots correspond to the cases $A=b$ (blue), $A>b$ (purple), and $A<b$ (red).}
    \label{Fig-R(r)_Vulcao}
\end{figure}
In Figure \ref{Fig-R(r)_Vulcao}, it is observed that the curvature scalar of the volcano-like surface exhibits, near \(r \approx 0\), a positive peak followed by a dip. For large values of \(r\), \(R(r) \to 0\), as expected, since the bump becomes negligible far from the crater. The presence of this maximum and the subsequent minimum is a consequence of the inflection points of the profile \(z(r)\), where the concavity of the surface changes sign. 

 As in the Gaussian case, $f_{\text{\tiny V}}(r)$ is positive definite and 
\begin{equation}
f_{\text{\tiny V}}^{\prime}(r) = -4r\left( \dfrac{2b^2}{b^2 - 2r^2}  + \dfrac{1}{b^2}\right)f_{\text{\tiny V}}(r)
\end{equation}
is not singular at $r = \sqrt{b/2}$, since $\displaystyle \lim_{r \rightarrow \sqrt{b/2}} f_{\text{\tiny V}}^{\prime}(r) = 0$. Thus, the scalar curvature of the volcano-shaped surface is also regular for all $r>0$.


\section{Dynamics of Fermions in Curved Geometries}
\label{sec:ryd}

In this section, we start from the eigenvalue equation associated with the Dirac Hamiltonian in curved spacetime, which reads
\begin{equation}
	\hat{H}_{C} \Psi(r, \theta) = E \Psi(r, \theta)  \label{E}
\end{equation}
to explore how the curvature of the surface modifies the behavior of electronic states in a graphene sheet. By writing the problem in a matrix form and adopting the separation of variables for the angular dependence with solutions of the form $e^{im\theta}$, we have
\begin{equation}
	\begin{pmatrix}
		0 & -i\hat{\Lambda} \\
		-i\hat{\Lambda}^\dagger & 0
	\end{pmatrix}
	\begin{pmatrix}
		\psi_A e^{im\theta} \\
		\psi_B e^{im\theta}
	\end{pmatrix}
	= \frac{E}{\hbar v_F}
	\begin{pmatrix}
		\psi_A e^{im\theta} \\
		\psi_B e^{im\theta}
	\end{pmatrix},
\end{equation}
where the operators $\hat{\Lambda}$ and $\hat{\Lambda}^\dagger$ are defined by
\begin{equation}
\hat{\Lambda} =	F(r)\partial_r + \frac{i}{r}\partial_\theta + A_\theta
\end{equation}
and
\begin{equation}
\hat{\Lambda}^\dagger = F(r)\partial_r - \frac{i}{r}\partial_\theta + A_\theta \,  .
\end{equation}

Thus, by applying these operators, we obtain the following ystem of coupled equations between the spinor components $\psi_A$ and $\psi_B$
\begin{equation}
	\begin{cases}
		-i\hat{\Lambda} (\psi_B e^{im\theta}) = \dfrac{E}{\hbar v_{\text{\tiny F}}} \psi_A e^{im\theta} \\
		-i\hat{\Lambda}^\dagger (\psi_A e^{im\theta}) = \dfrac{E}{\hbar v_{\text{\tiny F}}} \psi_B e^{im\theta},
	\end{cases} 
\end{equation}
Note the interdependence between the sub-lattices A and B.
 
Due to the axial symmetry of the Hamiltonian, only the \(z\)-component of the total angular momentum,
 \begin{equation}
     \hat J_z = -\,i\hbar\frac{\partial}{\partial\theta} + \frac{\hbar}{2}\,\sigma_z\, ,
 \end{equation}
is conserved, i.e.\ \([\hat H_C, \ \hat J_z]=0\). Consequently, we may choose common eigenstates of \(\hat H_C\) and \(\hat J_z\) with eigenvalue \(m\hbar\), which implies that the angular dependence of the spinor must appear as a phase factor \(e^{im\theta}\). Thus, we can write
\begin{equation}
\Psi(r,\theta) = e^{im\theta}
\begin{pmatrix}\psi_A(r)\\\psi_B(r)\end{pmatrix},
\end{equation}
where \(\psi_{A,B}(r)\) describes the radial behavior. In this way, the \(\theta\)-derivative disappears from the coupled equations and the problem reduces to an effectively one-dimensional system in \(r\), simplifying both analytical and numerical treatments.

 Proceeding with the decoupling, we obtain the following differential equations for $\psi_A$ and $\psi_B$:
\begin{equation}
\begin{aligned}
	F^2(r)\psi_A'' + F(r) [F'(r) + 2A_\theta]\psi_A' 
	&+ \big[F(r)A'_\theta \\-  \frac{m}{r^2}(m + F(r))
	&+ A_\theta^2\big]\psi_A = -\kappa^2 \psi_A \, ,
\end{aligned}
\label{Eq-Psi_A}
\end{equation}
\begin{equation}
\begin{aligned}
	F^2(r)\psi_B'' + F(r) [F'(r) + 2A_\theta]\psi_B' 
	&+ \big[F(r)A'_\theta \\-  \frac{m}{r^2}(m - F(r))
	&+ A_\theta^2\big]\psi_B = -\kappa^2 \psi_B\, ,
\end{aligned}
\label{Eq-Psi_B}
\end{equation}
where we have used here the linear dispersion relation of the electrons in a graphene sheet, $E = \pm \hbar v_{\text{\tiny F}} \kappa$ \cite{katsnelson2020physics}.

Here, we can see a symmetric behavior of the spinor components $\psi_{A}$ and $\psi_{B}$: by performing the transformation $m \rightarrow -m$ in Eq. (\ref{Eq-Psi_A}), we obtain Eq. (\ref{Eq-Psi_B}), so that
\begin{equation}
\psi_B \equiv \psi_A(-m) \, .
\end{equation}
Therefore, the behavior of a spinor component in different sub-lattices (A or B) is equivalent to an inversion of the angular momentum quantum number 
 $m$. Such a feature will be useful to reduce the effort of numerical implementations.

As a first approximation neglecting the variation of the effective Fermi velocity \cite{de2007charge} with respect to $A_\theta$, such that $F(r)^2 \approx 1$, we perform a transformation of the wave function as
\begin{equation}
	\psi_{\text{\tiny A,B}} = \mu(r) \chi_{\text{\tiny A,B}} \, ,
\end{equation}
in such a way that the Dirac equation yields decoupled equations for $\chi_A$ and $\chi_B$ in a form similar to the Klein-Gordon equation:
\begin{equation}
	-\dfrac{d^2\chi_{\text{\tiny A,B}}}{dr^2} +U^2_{\text{(eff)A,B}} \chi_{\text{\tiny A,B}}(r) = k^2 \chi_{\text{\tiny A,B}}(r)\, ,
\end{equation}
where
\begin{equation}
\begin{aligned}
U^2_{\text{(eff)}A,B} =\ & -FA'_\theta - A_\theta^2 + \frac{m}{r^2} \left( m \pm F \right) \\
& + \frac{1}{2} \frac{d}{dr} \Big[F( F'+ 2A'_\theta )\Big]
+ \frac{ 1}{4}F^2 \Big[ F' + 2A_\theta \Big]^2  ,
\end{aligned} \label{UAB2}
\end{equation}
is an effective potential and $\mu(r)$, defined as 
\begin{equation}
	\mu(r) = e^{-\int_{0}^{r} A_\theta(r') \, dr'} ,
\end{equation}
represents a geometric phase that modifies the electron wave function induced by the curvature of the graphene sheet. Consequently, the curved graphene sheet modifies the phase of $\psi_A$ and $\psi_B$, resulting in a geometric analog of the Aharonov–Bohm effect \cite{silva2024strain}. 

Similarly to the Eqs. (\ref{Eq-Psi_A}) and (\ref{Eq-Psi_B}),  we have a symmetry in the effective potential with respect to the sublattice and the spin:
\begin{equation}
U^2_{\text{(eff)}B} \equiv  U^2_{\text{(eff)}A}(-m) \, .
\end{equation}

\subsection{Fermion dynamics in the Gaussian surface}

Figure \ref{Fig-Atheta_Gaussina} shows the gauge pseudopotential $A_\theta$ given by Eq. (\ref{Atheta}) for the Gaussian surface in Figure \ref{Fig-Atheta_Gaussina}. It can be seen that when the amplitude of the Gaussian is greater than its width, the localized curvature becomes more intense, resulting in a stronger pseudomagnetic field which implies more pronounced effects on the dynamics of fermions. On the other hand, when the width exceeds the amplitude, the curvature is more distributed, and the pseudomagnetic field is reduced, reducing the geometric effects on the system. This demonstrates that the intensity of the pseudomagnetic field is a reflection of the localized curvature, directly impacting the structure of electronic states.
\begin{figure}[h!]
    \centering
    \includegraphics[width=0.9\linewidth]{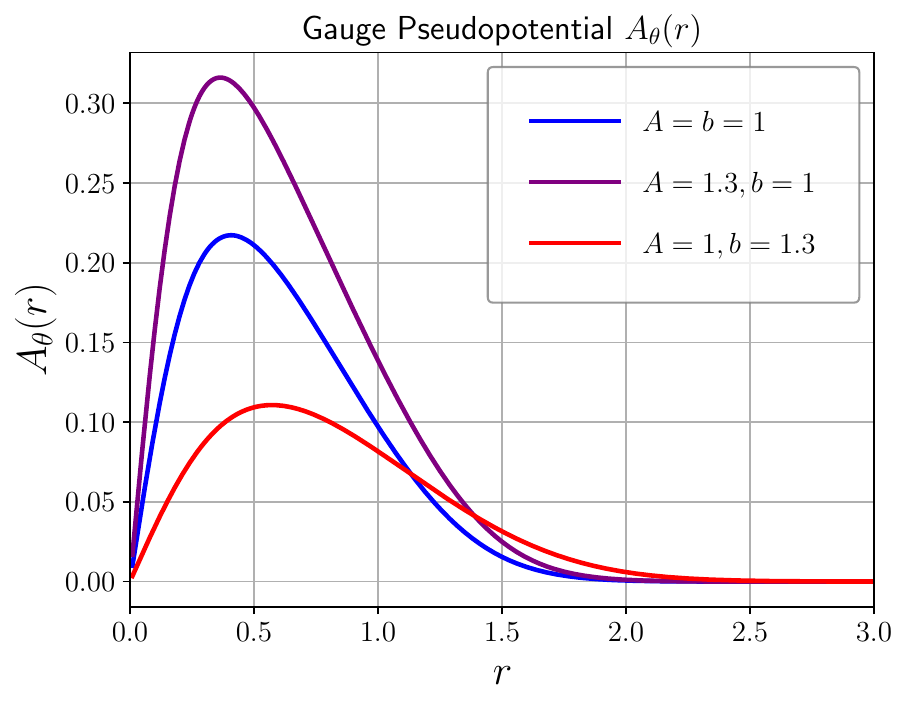}
    \caption{Plot of the gauge pseudopotential for the Gaussian surface for different $\alpha$ ratios: $A=b$ (blue), $A>b$ (purple), and $A<b$ (red).}
    \label{Fig-Atheta_Gaussina}
\end{figure}

In the absence of an external magnetic field (\( B = 0 \)), the behavior of the effective potentials reduces to
\begin{equation}
U^2_{\text{(eff)}A,B} = \mp \frac{m}{r} + A_\theta
\label{Eq-Potencial_Efetivo_B0}
\end{equation}   
and is essentially dominated by the term \( \pm \frac{m}{r} \), as can be seen in Figure \ref{Fig-U_eff_Gaussina_B0}, regardless of the specific form of the gauge pseudo-potential \( A_\theta(r) \) or the geometry of the surface, whether Gaussian or volcano-like. This is because, near the origin (\( r \to 0 \)), the  \( 1/r \) term diverges, while \( A_\theta(r) \) for Gaussian surfaces smoothly tends to zero, and even in cases where it diverges, such as in the volcano-like type, its contribution is suppressed compared with the universal behavior of the \( 1/r \) term. Physically, this effective potential represents the interaction between the fermion's orbital angular momentum and the local curvature of the surface, establishing either barriers or wells near the origin, depending on the sign of the quantum number \( m \). This contrasts significantly with the effective potential obtained by decoupling the Dirac equations.
 
\begin{figure}[h!]
    \centering
    \includegraphics[width=0.9\linewidth]{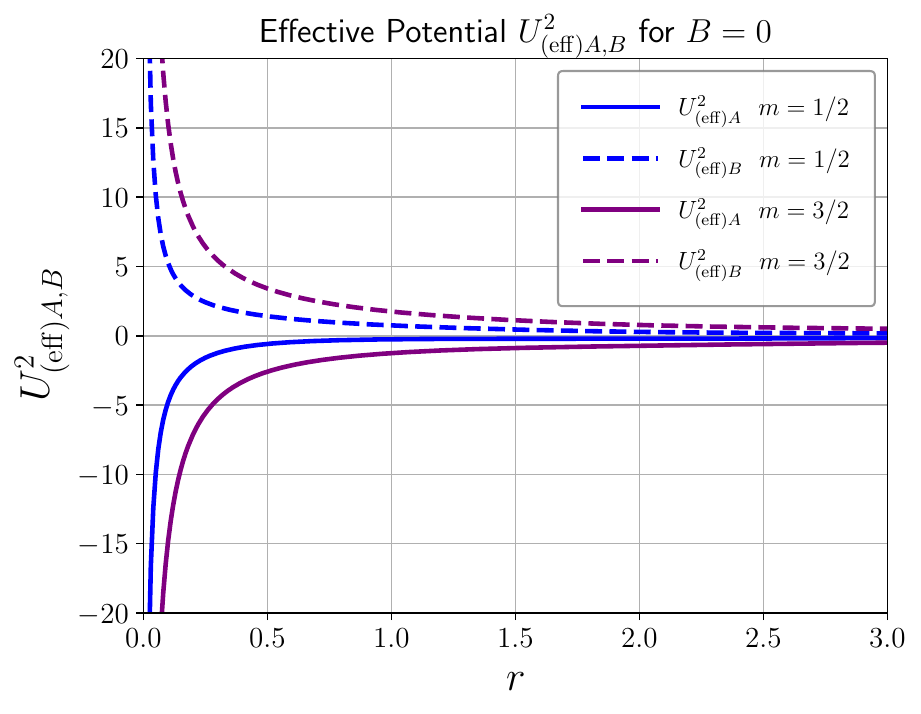}
    \caption{Plot of the effective potential $U^2_{\text{(eff)}A,B}$ in the absence of an external magnetic field (\( B = 0 \)) given by Eq. (\ref{Eq-Potencial_Efetivo_B0}) for the Gaussian surface with $A = 1$ and $b = 1$. The thick lines correspond to $U^2_{\text{(eff)}A}$, while the dashed ones, to $U^2_{\text{(eff)}B}$.}
    \label{Fig-U_eff_Gaussina_B0}
\end{figure}


 The figure \ref{UAg}  illustrates the behavior of the effective potential $U^2_{\text{(eff)}A,B}$ given by Eq. (\ref{UAB2}) for different values of $m$. It is observed that as $ r \to 0$, the dominant term, which directly depends on $m$, diverges and surpasses the contribution of the other terms. The remaining terms, in turn, tend to remain regular in this region, not significantly altering the behavior of the dominant term proportional to $1/r^2$. This divergence, responsible for the local potential barrier or well, can be attractive or repulsive depending on the sign of $m$. 

 A noteworthy result is the the appearance of a local minimum in the effective potential which indicates metastable states, instead of a purely repulsive barrier as in the case without magnetic field. We also study the effective potential for different values of $A$ and $b$. For $A>b$, the central well of $U^2_{\text{(eff)}A,B}$ appears more prominent, which leads us to interpret that: when the Gaussian surface becomes higher, the electron interacts more strongly with the graphene sheet. On the other hand, for $A< b$ (a low and spread surface), the potential well tends to diminish and the effective potential approaches the case without a magnetic field.  
\begin{figure}[t!]
    \centering
    \includegraphics[width=0.9\linewidth]{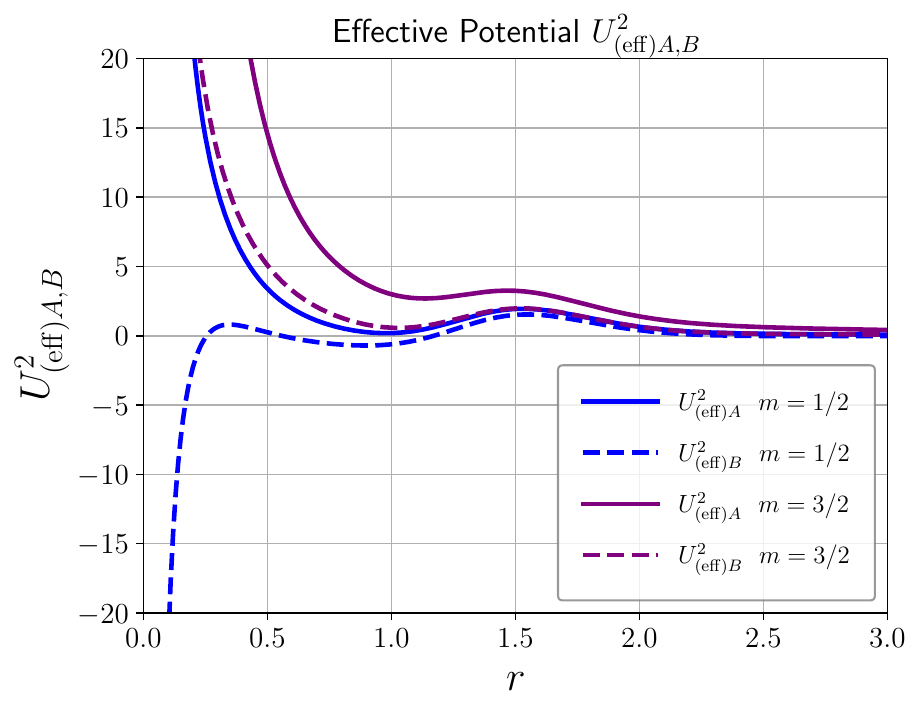}
     \caption{Plot of the effective potential for $U^2_{\text{(eff)}A,B}$ given by Eq. (\ref{UAB2}) for the Gaussian surface with $A = 1$ and $b = 1$. The thick lines corresponds to $U^2_{\text{(eff)}A}$, while the dashed one, to $U^2_{\text{(eff)}B}$.}
\label{UAg}
\end{figure}

Given that experiments show that the Gaussian width is greater than its height, we can expand the function $F(r)$ in a Taylor series up to first order around $\alpha=0$. Neglecting the variation of the effective Fermi velocity relative to $A_\theta$ and making such approximations, we have
\begin{equation}
	-\chi_{A,B}''(r) + \frac{m}{r^2} (m\pm1) \chi_{A,B}(r) = \kappa^2 \chi_{A,B}(r)\, . \label{chi}
\end{equation}
These differential equations yield solutions as linear combinations of Bessel functions of the first $J_{\nu}$ and second $Y_{\nu}$ kinds in the form:
\begin{equation}
   \psi_{A,B}=\text{exp}\left( {\frac{\alpha }{4}e^{-\frac{2r^2}{b^2}}}\right) \sqrt{r} \left[ C_1 \, J_{\frac{1 \pm 2m}{2}}(\kappa r) + C_2 \, Y_{\frac{1 \pm 2m}{2}}(\kappa r) \right].
\end{equation}
Due to the divergence of the Bessel function of the second kind at the origin, we impose that $C_2=0$. Therefore, the general solution is given by
 \begin{equation}
   \psi_{A,B}=C_1\, \text{exp}\left({\frac{\alpha }{4}e^{-\frac{2r^2}{b^2}}}\right) \sqrt{r}   \, J_{\frac{1 \pm 2m}{2}}(\kappa r) \, .
   \label{Eq-Psi_AB_sem_Fermi}
\end{equation}
These solutions can be normalized using
	\begin{equation}
		\int_0^{2\pi} d\theta \int_0^\infty dr\, r\, \rho(r,\theta) = 1\, ,
	\end{equation}
where
	\begin{equation}
		\rho(r,\theta)=\Psi^\dagger(r,\theta)\Psi(r,\theta)= |\psi_A(r)|^2+|\psi_B(r)|^2.
	\end{equation}   
	
The spinor components, $\psi_A$ and $\psi_B$, represent the probability amplitudes associated with the A and B sublattices of graphene, respectively. Thus, $|\psi_A|^2$ and $|\psi_B|^2$ provide the probability density of finding the electron in each of these sublattices in a given configuration \cite{katsnelson2020physics}.

In Figures \ref{psiAgg} and \ref{psiBgg}, we can observe the probability distribution for the components $\psi_A$ and $\psi_B$ as a function of the parameter $m$. When $m=1/2$, the component $\psi_B$ dominates, exhibiting a significantly higher probability density in sublattice B  near the origin, where the Gaussian bump is located; in contrast, the component $\psi_A$ has lower amplitude near the Gaussian bump and behaves like a plane wave. This suggests that the electron tends to behave as a free particle in the sublattice $A$.  For $m = 3/2$, the peaks in the wave functions get lower, especially for $\psi_A(r)$, and the solutions perform as plane waves. Another interesting result is that the wave functions begin its oscillation furthest from the origin, revealing a classically forbidden region.  Thus, the electron with $m = 3/2$ moves away from the Gaussian surface. We have also looked at $m = 5/2$ and such separation became even more evident. We concluded that electrons with higher $m$ propagate away from the Gaussian surface.

\begin{figure}[h!]
    \centering
    \includegraphics[width=0.9\linewidth]{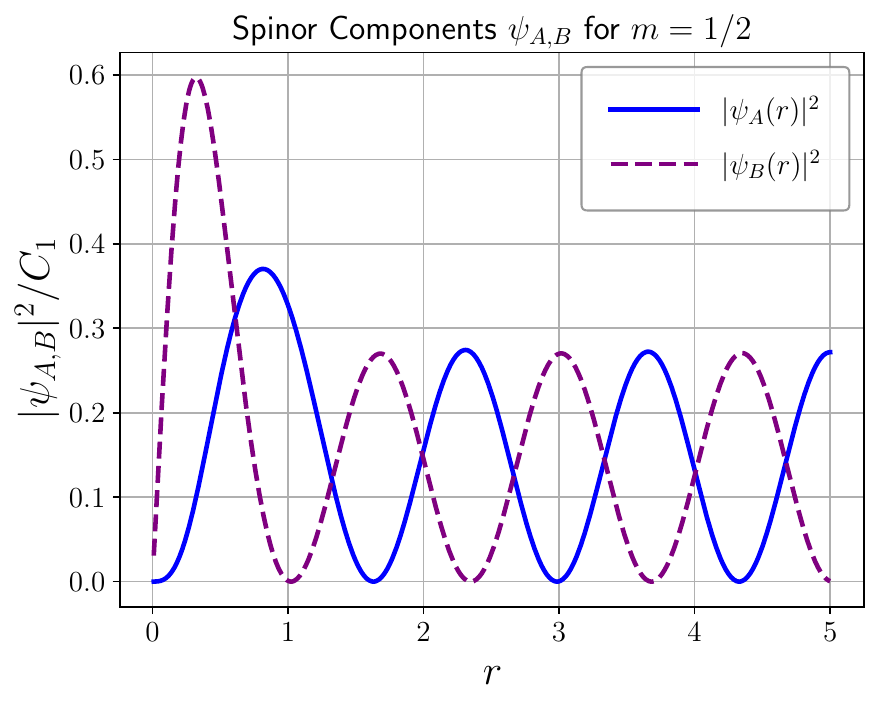}
    \caption{Probability density of $\psi_{A}(r)$ and $\psi_{B}(r)$ for $m = 1/2$ neglecting variations of the Fermi velocity, given by Eq. (\ref{Eq-Psi_AB_sem_Fermi}). Here we have set $A = 1.5$, $b = 1.0$ and $\kappa = 2.35$.}
    \label{psiAgg}
\end{figure}
\begin{figure}[h!]
    \centering
\includegraphics[width=0.9\linewidth]{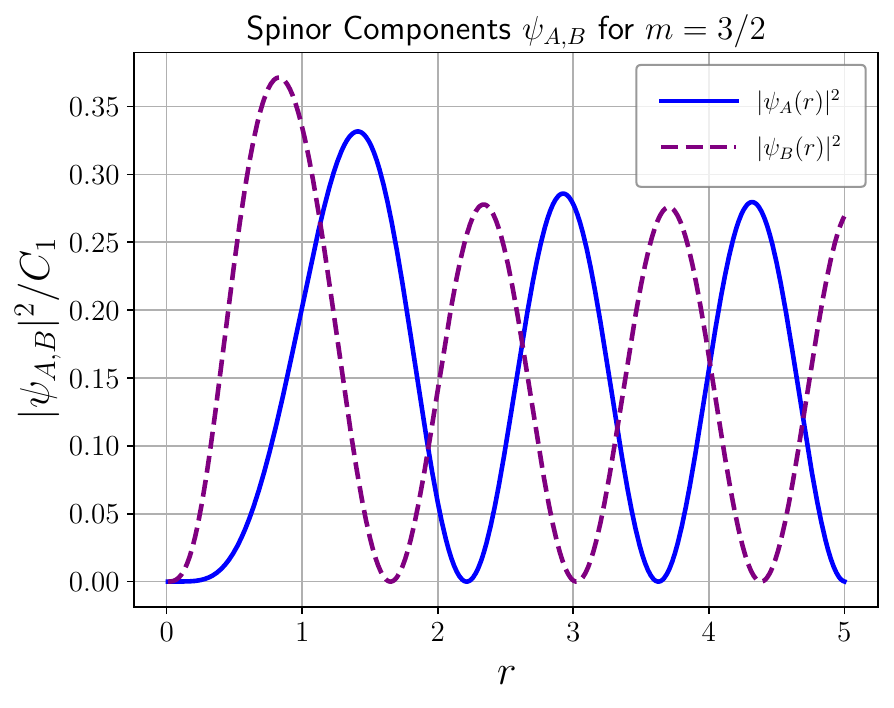}
    \caption{Probability density of $\psi_{A}(r)$ and $\psi_{B}(r)$ for $m = 3/2$  neglecting variations of the Fermi velocity, given by Eq. (\ref{Eq-Psi_AB_sem_Fermi}). We used the same parameters of the previous case.}
    \label{psiBgg}
\end{figure}

Furthermore, as $r$ increases, both wave functions exhibit an oscillatory behavior typical of a free particle, reflecting the absence of confinement at long distances. This asymptotic regime confirms that, for large values of $r$, the solution describes plane waves, as expected for scattering in a medium without significant confining potentials.
We observe  that the probability distributions are closely related to the behavior of the effective potential. Consequently, the probability of finding the electron in each sublattice reflects the interaction between these potentials, leading to different configurations of $\psi_A$ and $\psi_B$ as the quantum number $m$ varies. These observations highlight the importance of the spin quantum number $m$ in the dynamics of electrons in the graphene sheet in the presence of localized curvature.

To analyze the probability density including the variation of the Fermi velocity, we performed a numerical treatment of the Eqs. (\ref{Eq-Psi_A}) and (\ref{Eq-Psi_B}), but first we will present the same analysis of this subsection for the volcano-shaped surface.


\subsection{Fermion dynamics in the volcano-like surface}

Following the same procedure as in  the previous subsection, we plot  the pseudopotential  $A_\theta$ for the volcano-shaped surface, given by Eq. (\ref{Atheta}), in Figure  \ref{athetav}.  Unlike the Gaussian case, where $A_\theta$ tends to zero as  $r \to 0$, the behavior of the gauge pseudopotential for the volcano-shaped surface diverges near the origin.  This behavior suggests that the local curvature generates an effective field that is highly concentrated near $r = 0$. We emphasize that the physically relevant regime for this work corresponds to $ b > A$; therefore, all subsequent results will be presented under this condition.
\begin{figure}[h!]
    \centering
    \includegraphics[width=0.9\linewidth]{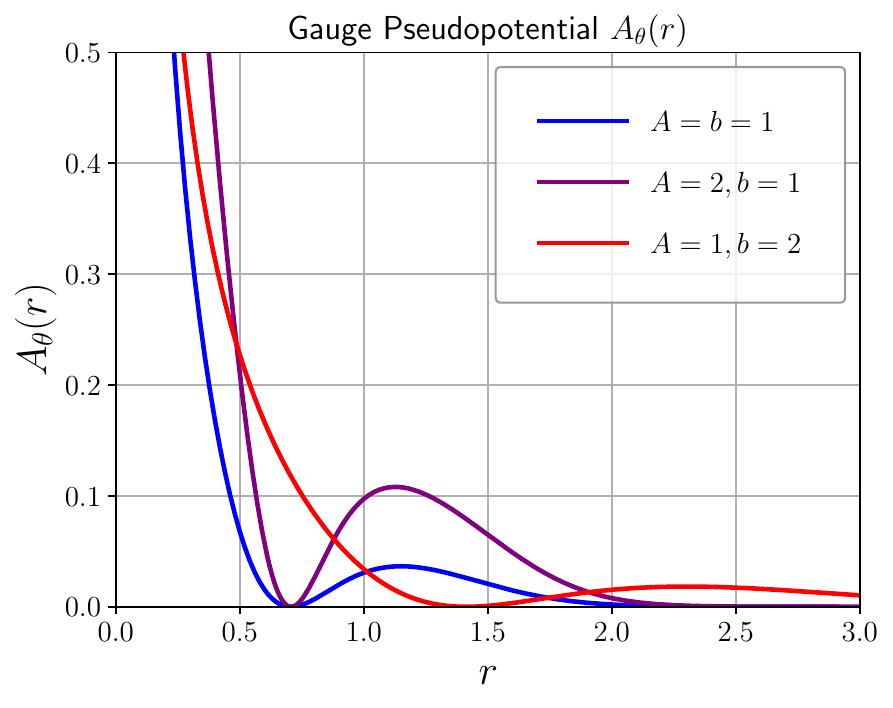}
    \caption{Plot of the gauge pseudopotential for the volcano-shaped surface for different $\alpha$ ratios: $A=b$ (blue), $A>b$ (purple), and $A<b$ (red).}
    \label{athetav}
\end{figure}

Again, we obtain a pair of decoupled equations in a form similar to Klein-Gordon equation (\ref{chi}), whose general solutions for  $\psi_A$  and  $\psi_B$ are
\begin{equation}
   \psi_{A,B}(r)=C\, e^{-[\frac{1}{4}\ln r]} \, \sqrt{r} \, J_{\frac{1 \pm 2m}{2}}(k r) \, .
   \label{Eq-Psi_SemFermi_Vulcao}
\end{equation}
\begin{figure}[h!]
    \centering
    \includegraphics[width=0.9\linewidth]{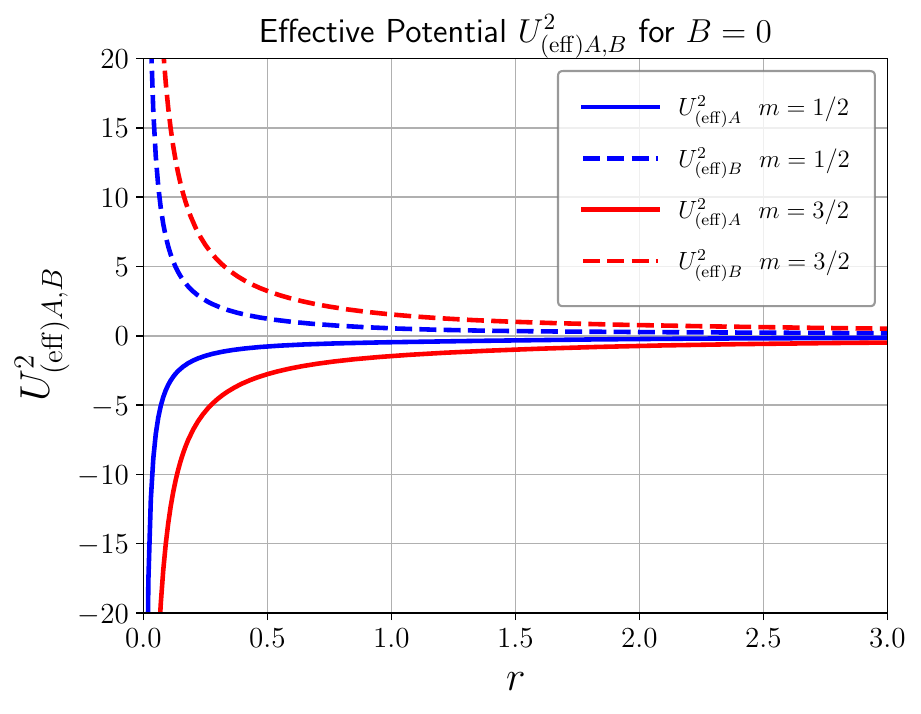}
    \caption{Plot of the effective potential for $U^2_{\text{(eff)}A,B}$ in the absence of an external magnetic field (\( B = 0 \)) given by Eq. (\ref{Eq-Potencial_Efetivo_B0}) for the volcano-shaped surface with $A = 1$ and $b = 2$. The thick lines corresponds to $U^2_{\text{(eff)}A}$, while the dashed ones, to $U^2_{\text{(eff)}B}$.}
    \label{Fig-U_eff_Vulcao_B0}
\end{figure}
\begin{figure}[h!]
    \centering
    \includegraphics[width=0.9\linewidth]{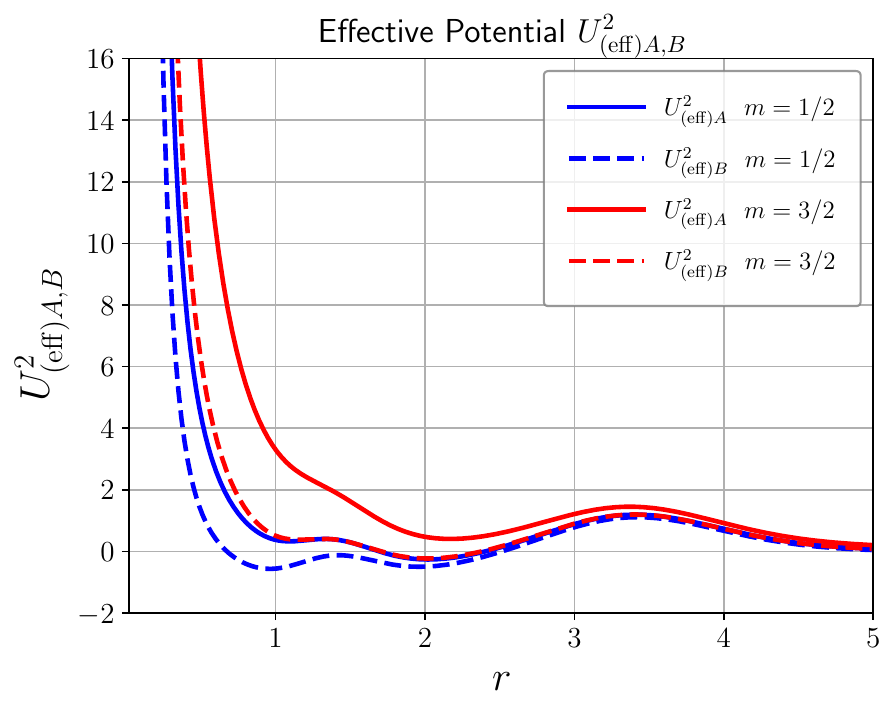}
    \caption{Plot of the effective potential for $U^2_{\text{(eff)}A,B}$ given by Eq. (\ref{UAB2}) for the volcano-shaped surface with $A = 1$ and $b = 2$. The thick lines corresponds to $U^2_{\text{(eff)}A}$, while the dashed one, to $U^2_{\text{(eff)}B}$.}
    \label{UB2V}
\end{figure}

We also plot the effective potentials given by Eq. (\ref{UAB2}) and Eq. (\ref{Eq-Potencial_Efetivo_B0}) (without magnetic external field) in  Figures \ref{Fig-U_eff_Vulcao_B0} and \ref{UB2V}, respectively, for the volcano-shaped bump.

The effective potentials for the volcano-type surface exhibit in all cases a divergent behavior as $r \to 0$. This result occurs due to the dominant term, which is proportional to $m/r^2$, causing the surface curvature to play only a secondary role near the origin. When the magnetic field is turned on, the effective potential given in Eq. (\ref{UAB2}) presents small local wells and barriers indicating interactions between the electron and the graphene sublattices, which may induce metastable states. It is interesting to note that the potential well is more pronounced for spin-
$1/2$ electrons in the B sublattice.
By analyzing the probability densities for the two sublattices $|\psi_A|^2$ and $|\psi_B|^2$  on the volcano-type surface in comparison to the Gaussian surface, as shown in Figures \ref{psiAV2} and \ref{psiBV2}, it becomes clear that the sub-lattice B is energetically more favored. Although asymptotically, the wave functions behave as plane waves which indicate free electrons, the wavefunction $\psi$ has higher peaks in the B sub-lattice in the volcano bump. The main difference between the volcano and the Gaussian lies in the geometric phase $\mu(r)$ associated with the pseudopotential $A_\theta$, which modifies the functions $ \psi_A$ and $ \psi_B$. For both surfaces, in the absence of an external field, we note that $ \psi_B$ can be obtained from $ \psi_A$ by swapping $ m \rightarrow -m $, highlighting an intrinsic interdependence between the two components of the spinor. 
\begin{figure}[h!]
    \centering
    \includegraphics[width=0.9\linewidth]{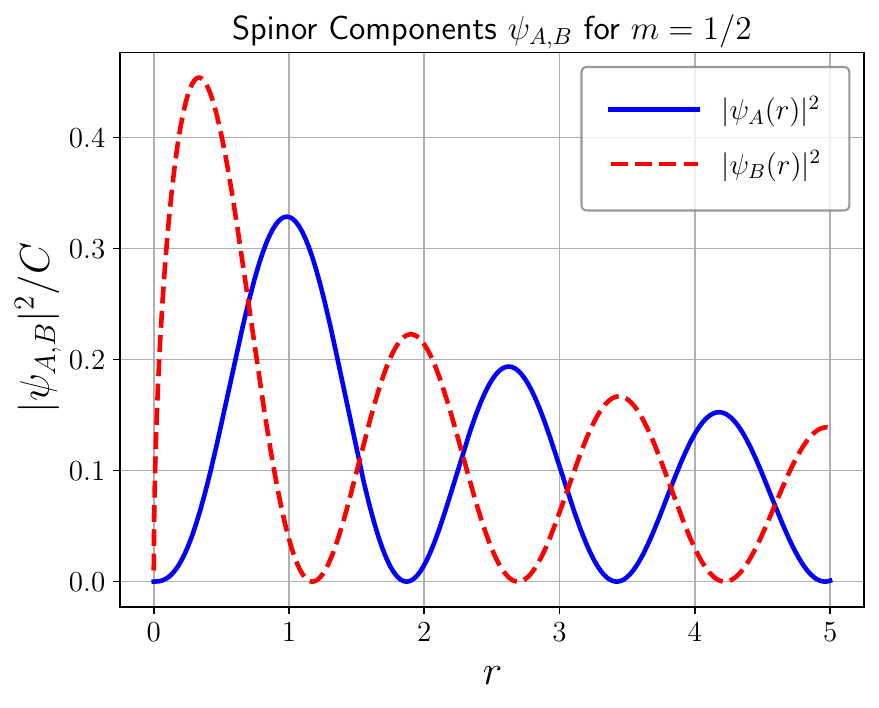}
    \caption{Probability density of $\psi_{A}(r)$ and $\psi_{B}(r)$ for $m = 1/2$ neglecting variations of the Fermi velocity, given by Eq. (\ref{Eq-Psi_SemFermi_Vulcao}) for the Volcano-shaped bump. Here we have set $A = 1.0$ and $b = 2.0$. The scale argument in  the Bessel Function was set to $k = 2.0	$.}
    \label{psiAV2}
\end{figure}
\begin{figure}[h!]
    \centering
    \includegraphics[width=0.9\linewidth]{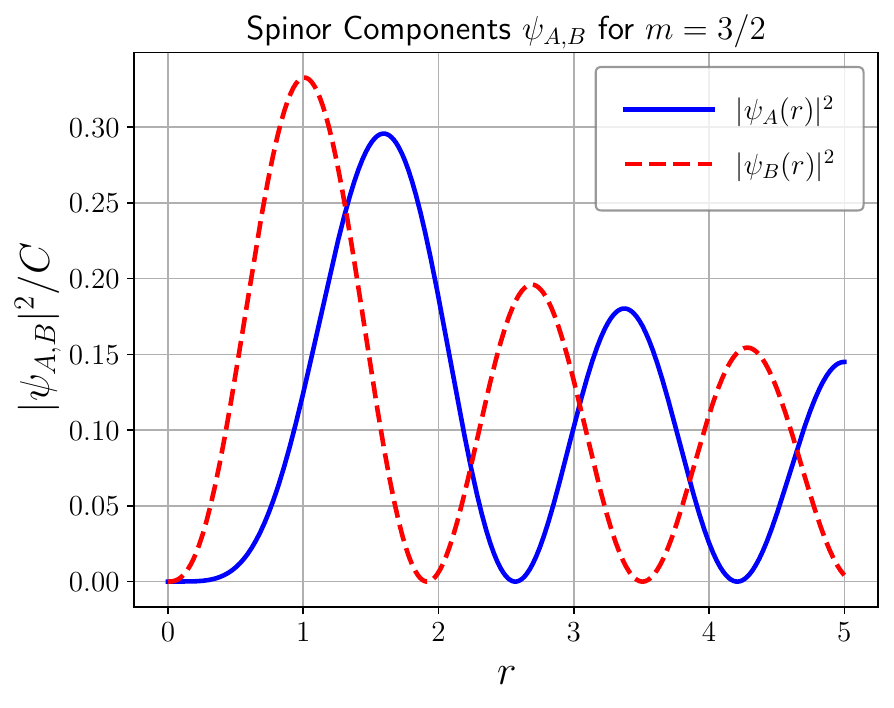}
\caption{Probability density of $\psi_{A}(r)$ and $\psi_{B}(r)$ for $m = 3/2$  neglecting variations of the Fermi velocity, given by Eq. (\ref{Eq-Psi_SemFermi_Vulcao}). We used the same parameters of the previous case.}
    \label{psiBV2}
\end{figure}


\section{Numerical Analysis}
\label{Sec-Numerical}
In order to study the physical behavior of electrons in graphene lattices with Gaussian and volcano-shaped bumps, taking into account variations in the Fermi velocity, we numerically analyzed Eqs. (\ref{Eq-Psi_A}) and (\ref{Eq-Psi_B}). The resulting differential equations are formulated as Sturm–Liouville problems. Due to the nontrivial forms of the coefficients in such boundary value problems, a numerical treatment is necessary. Thus, we employed a finite difference scheme with second-order truncation error, where the derivatives of $\psi_{\text{\tiny{A,B}}}$ were approximated using centered finite differences. This approach is known as the matrix method. \cite{Matrix_Method}. This type of equations also appears in High Energy Physics in the context of braneworld model with codimension-2 with fermion localization procedures \cite{Fermions_Charuto}, and the matrix method was also applied in the same context for gravity and gauge fields localization \cite{Graviton_Charuto, Graviton_Julio, Gauge_Charuto}.

The numerical solution of a Sturm–Liouville problem provides both the eigenvalues (which correspond to the electron energy levels $\kappa$) and the corresponding eigenfunctions, which are the wave functions $\psi_{\text{A,B}}$. The boundary conditions were
\begin{equation}
\psi_{\text{A,B}}(0) = 0 \qquad \text{and} \qquad \psi^{\prime}_{\text{A,B}}(+\infty) = 0 \, .
\label{Eq-Cond_Contorno}
\end{equation}
Note that these conditions are also satisfied in the case where variations in the Fermi velocity are neglected as can be seen in Figures \ref{psiAgg}, \ref{psiBgg}, \ref{Fig-U_eff_Vulcao_B0} and \ref{UB2V}.

 Firstly, we discretized the domain $r \in [0.01 \, , \, 5.00]$ with step size $h = 0.001$, resulting  in a $5000\times5000$ tridiagonal linear system. In Figure \ref{Fig-Epectro} we plot the first energy eigenvalues $\kappa_n$, where $n$ is integer, for the Gaussian bump case. Note that, because the problem of fermion propagation in graphene is placed in a finite domain in the numerical treatment, this leads to the quantization of the energy of these fermions. The energy spectrum is real and linearly increasing and spin-$3/2$ fermions carry slightly higher energies than spin-$1/2$ fermions. Variations in the geometry of the Gaussian surface (changes in parameters $A$ and $b$) did not lead to significant differences in the results for the energy spectrum. Moreover, we found very similar characteristics for the volcano-shaped bump. 
\begin{figure}[h!]
    \centering
    \includegraphics[width=0.9\linewidth]{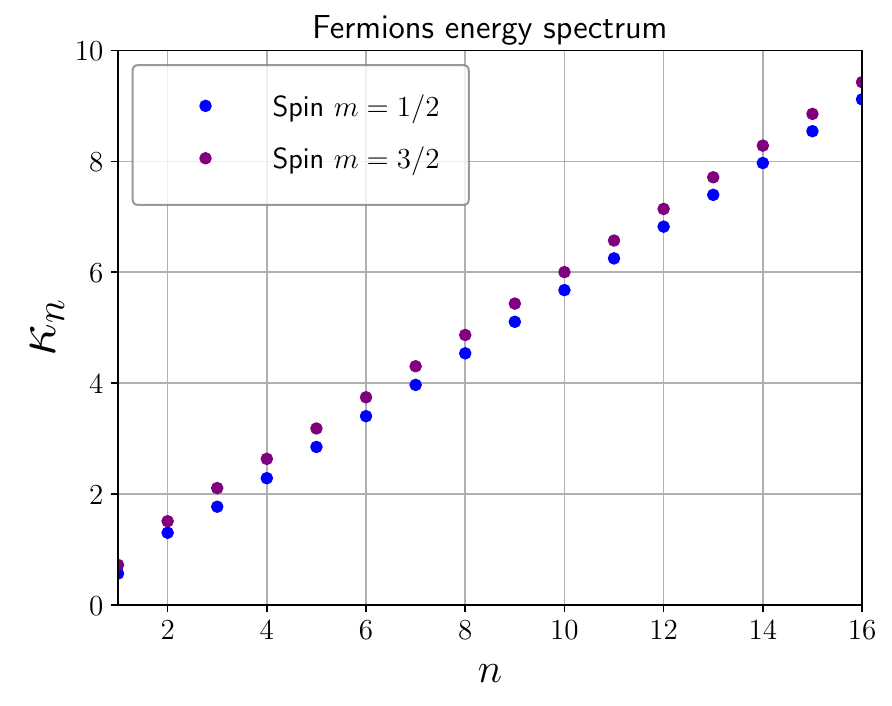}
    \caption{Plot of the fermions energy spectrum $\kappa_n$ for the Gaussian surface with $a = 1.3$ and $b = 1.0$.}
    \label{Fig-Epectro}
\end{figure}
  
Distinct and particularly interesting results were observed in the analysis of the eigenfunctions. We plot in Figures \ref{Fig-Psi_Gaussian_1/2} and \ref{Fig-Psi_Gaussian_3/2} the fifth eigenfunction for the Gaussian surface with $m = 1/2$ and $m = 3/2$, respectively. First of all, for spin $1/2$ fermions, when considering Fermi velocity, such fermions interact more strongly with the B sublattice of graphene since their wave functions exhibit more prominent second peaks, which increases the probability density of finding the fermions in the Gaussian bump region. For spin-$3/2$ fermions, we can see that only the first cycle of their wave functions shows prominent peaks, indicating a smaller region with a higher probability of finding these electrons compared to spin 1/2 fermions. Furthermore, these functions move further away from the origin, while still maintaining the boundary conditions in the derivative of $\psi_{\text{A,B}}$.

\begin{figure}[h!]
    \centering
    \includegraphics[width=0.9\linewidth]{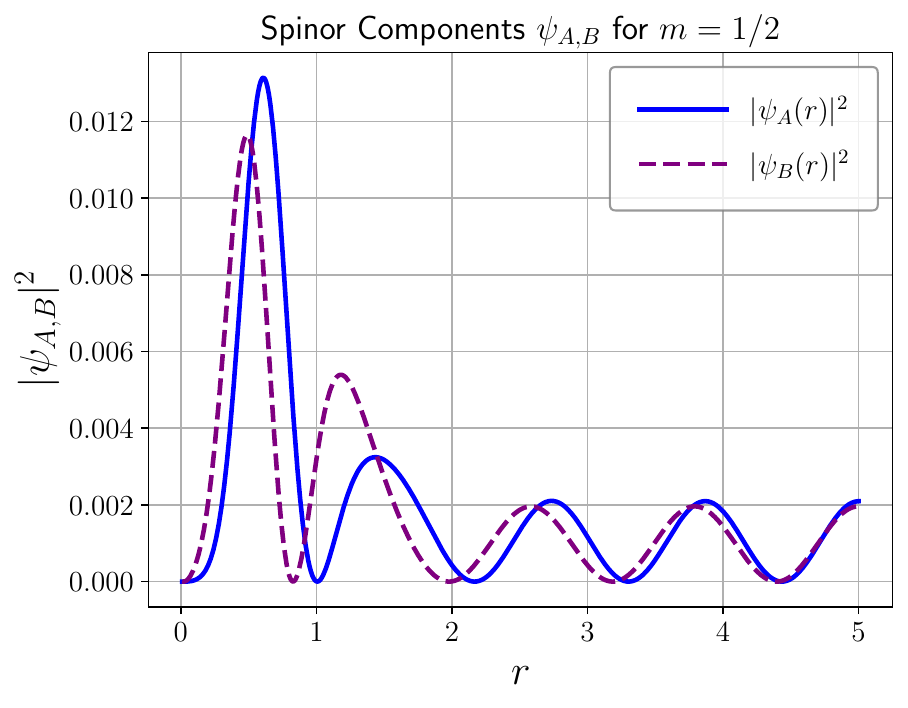}
\caption{ Numerical results for the probability density of $\psi_{A}(r)$ and $\psi_{B}(r)$ for $m = 1/2$ for the Gaussian surface considering variations of the Fermi velocity. Here we have set $A = 1.3$ and $b = 1.0$. The corresponding eigenfuctions have $\kappa_5 = 3.0081$.}
    \label{Fig-Psi_Gaussian_1/2}
\end{figure}

\begin{figure}[h!]
    \centering
    \includegraphics[width=0.9\linewidth]{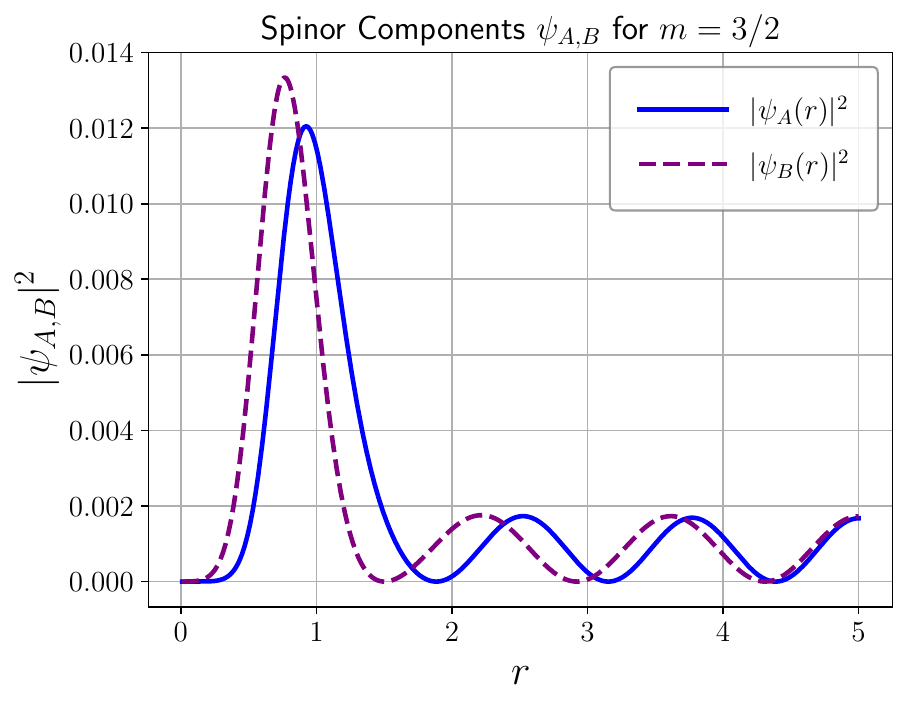}
\caption{ Numerical results for the probability density of $\psi_{A}(r)$ and $\psi_{B}(r)$ for $m = 3/2$ considering variations of the Fermi velocity. Here we have set $A = 1.3$ and $b = 1.0$. The corresponding eigenfuctions have $\kappa_5 =  3.2774$.}
    \label{Fig-Psi_Gaussian_3/2}

\end{figure}

Regarding the volcano-shaped surface, a noteworthy result emerges: the fermions have an increased probability density of being found further away from the origin, which can be interpreted as corresponding to the slope of the volcano. We illustrate this result in Figures \ref{Fig-Psi_Vulcao_1/2} and \ref{Fig-Psi_Vulcao_3/2}. In all cases, the eigenfunctions behave as plane waves asymptotically.

It is important to mention that numerical solutions of Sturm-Liouville problems in a finite differences-based scheme are good to approximate the first eigenvalues \cite{Matrix_Method}. However, these are properly the limits of physical interest. In fact, as the energy $\kappa$ becomes very hight, plane waves with high frequencies are obtained, as expected from the de continuum limit of the quantum mechanics. Furthermore, the analysis of the spectrum accuracy with respect to the discretization mesh can be verified as done in Reference \cite{Veras:2017nke}.

\begin{figure}[h!]
    \centering
    \includegraphics[width=0.9\linewidth]{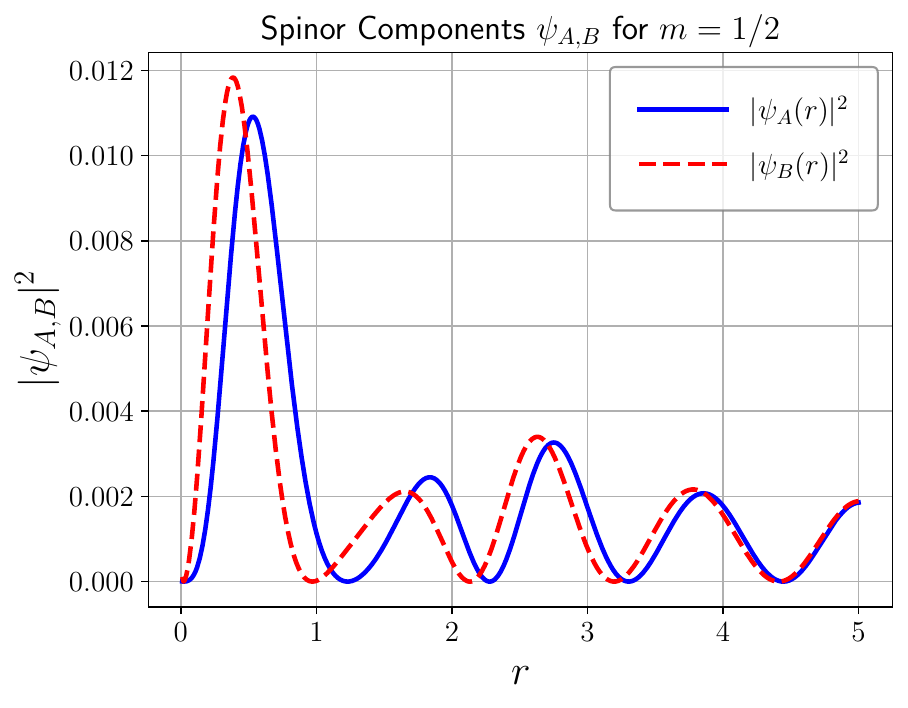}
\caption{ Numerical results for the probability density $|\psi_{A}(r)|^2$ and $|\psi_{B}(r)|^2$ for $m = 1/2$ for the Volcano-shaped surface considering variations of the Fermi velocity. Here we have set $A = 1.3$ and $b = 2.0$. The corresponding eigenfuctions have $\kappa_5 = 3.1595$.}
    \label{Fig-Psi_Vulcao_1/2}
\end{figure}
\begin{figure}[h!]
    \centering
    \includegraphics[width=0.9\linewidth]{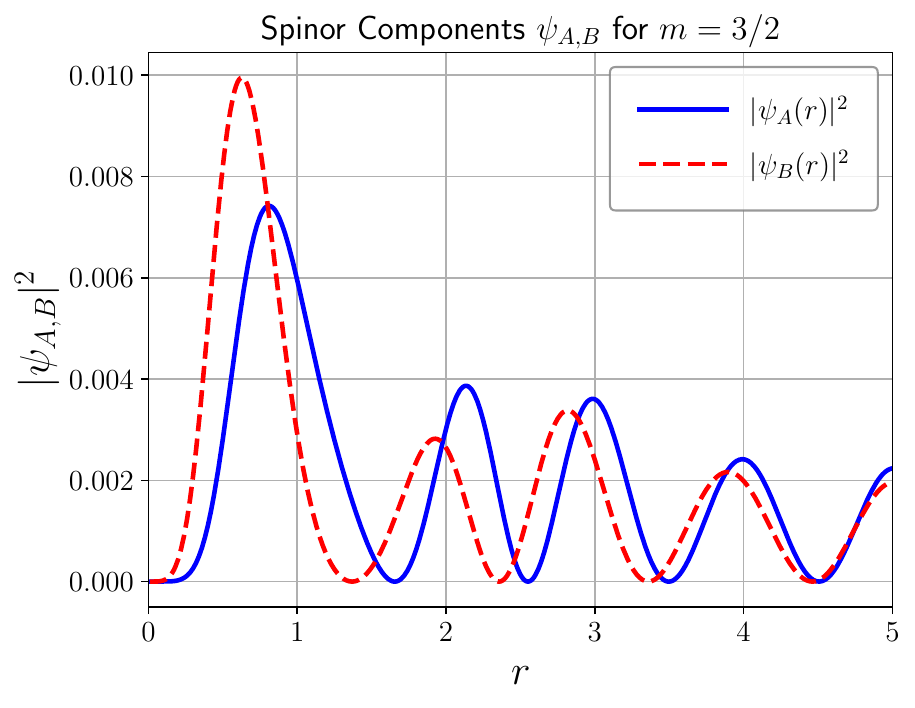}
\caption{ Numerical results for the probability density of $|\psi_{A}(r)|^2$ and $|\psi_{B}(r)|^2$ for $m = 3/2$ for the Volcano-shaped surface considering variations of the Fermi velocity. Here we have set $A = 1.3$ and $b = 2.0$. The corresponding eigenfuctions have $\kappa_5 = 3.7841$.}
    \label{Fig-Psi_Vulcao_3/2}

\end{figure}


\section{Concluding Remarks}
\label{sec:exp}

In this work, we investigate how localized curvature affects the dynamics of massless Dirac fermions in graphene sheets both in the absence and in the presence of an external magnetic field. By coupling the Dirac equation to curved surfaces with axial symmetry \cite{de2007charge}—one with a volcano-type profile and the other with a Gaussian profile—we find that the presence of curvature alters the electronic probability distribution.

To solve the problem, we express the solution of the Dirac equation as the product of two functions, $\psi = \mu \cdot \chi$, where $\mu(r)$ is a function that modifies the phase of the spinor components $\psi_{A,B}$ and depends exclusively on the curvature term $A_\theta(r)$. This dependence leads to the interpretation that curvature induces an effect analogous to the Aharonov–Bohm effect, but of purely geometric origin, acting as a phase potential that influences the propagation of fermions in the system \cite{furtado2008geometric,silva2024strain}.

When considering only the electron-curvature interaction, the probability densities indicate that electrons tend to accumulate in regions where the curvature is most intense \cite{de2007charge}. In the absence of an external field, we observe unbound states, where the energy can take continuous values. However, upon introducing a constant external magnetic field, we observe energy quantization into discrete levels, analogous to Landau levels, indicating the formation of bound states. A noteworthy result is the emergence of a local minimum in the effective potential which indicates metastable states, instead of purely repulsive barrier as in the case without magnetic field. Such a potential well becomes more pronounced as the height of the surface increases.

A remarkable aspect of the results is the dependence of the electronic distribution on the total angular momentum $m$ in the $z$ direction, which determines the preference for one of graphene’s sublattices. The spinor components, representing the probability amplitudes in the sublattices, exhibit asymmetric occupation depending on $m$, with one sublattice being significantly more populated than the other.

Additionally, we found that the shape of the total probability density $\rho(r)$ reflects the interaction between curvature, the magnetic field, and the effective potential: even though the curvature is maximal at the origin (in the case of the Gaussian surface), electrons never localize exactly at $r=0$ due to the divergence of the quadratic effective potential in that region. The resulting electronic distribution forms a radial ring around the origin, whose minimum radius decreases with increasing magnetic field strength but never reaches the center.

 We also noted a particular symmetry: the behavior of a spinor component in different sub-lattices (A or B) is equivalent to reversing the sign of its spin quantum number $m$. Furthermore, we numerically studied the electron energy spectrum and the corresponding eigenfunctions. 

Through a refined numerical treatment, we obtained the energy spectrum and wave functions of the fermions. The discretized energy levels exhibit linear behavior, and spin-$3/2$ fermions exhibit a slightly higher tower of energy states than spin-$1/2$ fermions. There was no significant difference between the two surfaces considered. Interesting features were observed in the corresponding wave functions. Asymptotically, the fermions behave as plane waves in all cases, as expected. However, in the vicinity of the considered geometries, there is a higher probability of finding fermions near the Gaussian and volcano-shaped surfaces, and specifically for the latter, there is a probability of finding these fermions in a slightly more distant region, which we can infer to be the slope of the Volcano. Thus, we conclude that the curved surfaces on the graphene sheet cause interactions with fermions, allowing the formation of bound states.

Overall, our results demonstrate that surface curvature can be leveraged in conjunction with external magnetic fields as an additional mechanism for controlling electronic localization in two-dimensional materials such as graphene.

\section{acknowledgments}

The authors thank the Conselho Nacional de Desenvolvimento Científico e Tecnológico (CNPq), for financial support, grant nº 304120/2021-9 (JEGS) and  grant nº 131749/2023-4 (ARNL) and Fundação Cearense de Apoio ao Desenvolvimento Científico e Tecnológico (FUNCAP) for also financial support, grant nº BP6-0241-00324.01.00/25 $-$ NUP 31052.002489/2025-39 (DFSV).
\newpage

\bibliographystyle{unsrt}
\bibliography{Referencias}

@article{geim2009graphene,
  author = {A. K. Geim},
  journal = {Science},
  volume = {324},
  pages = {1530--1534},
  year = {2009}
}

@article{seung1988defects,
  author    = {Hyunjune Sebastian Seung and David R. Nelson},
  journal   = {Phys. Rev. A},
  volume    = {38},
  pages     = {1005},
  year      = {1988}
}

@article{meyer2007structure,
  author    = {Jannik C. Meyer and Andre K. Geim and Mikhail I. Katsnelson and Konstantin S. Novoselov and Tim J. Booth and Siegmar Roth},
  journal   = {Nature},
  volume    = {446},
  pages     = {60},
  year      = {2007}
}

@article{meyer2007roughness,
  author    = {Jannik C. Meyer and A. K. Geim and M. I. Katsnelson and K. S. Novoselov and D. Obergfell and S. Roth and C. Girit and A. Zettl},
  journal   = {Solid State Commun.},
  volume    = {143},
  pages     = {101},
  year      = {2007}
}

@article{de2007charge,
  author    = {Fernando de Juan and Alberto Cortijo and Mar{\'\i}a A. H. Vozmediano},
  journal   = {Phys. Rev. B},
  volume    = {76},
  pages     = {165409},
  year      = {2007}
}

@article{sutter2009scanning,
  author    = {Eli Sutter and D. P. Acharya and J. T. Sadowski and P. Sutter},
  journal   = {Appl. Phys. Lett.},
  volume    = {94},
  year      = {2009}
}

@article{ishigami2007atomic,
  author    = {Masa Ishigami and Jyong-Hao Chen and William G. Cullen and Michael S. Fuhrer and Ellen D. Williams},
  journal   = {Nano Lett.},
  volume    = {7},
  pages     = {1643},
  year      = {2007}
}

@article{geim2007graphene,
  author    = {Andre Geim and K. Novoselov},
  journal   = {Nat. Mater.},
  volume    = {6},
  pages     = {10}, 
  year      = {2007}
}

@article{monteiro2023dirac,
  author    = {L. N. Monteiro and C. A. S. Almeida and J. E. G. Silva},
  journal   = {Phys. Rev. B},
  volume    = {108},
  pages     = {115436},
  year      = {2023}
}

@article{de2012space,
  author    = {Fernando de Juan and Mauricio Sturla and Mar{\'\i}a A. H. Vozmediano},
  journal   = {Phys. Rev. Lett.},
  volume    = {108},
  pages     = {227205},
  year      = {2012}
}

@article{furtado2008geometric,
  author={Furtado, Claudio and Moraes, Fernando and Carvalho, AM de M},
  journal={Phys. Lett. A},
  volume={372},
  number={32},
  year={2008},
  publisher={Elsevier}
}

@article{atanasov2015helicoidal,
  author={Atanasov, Victor and Saxena, Avadh},
  journal={Phys. Rev. B},
  volume={92},
  number={3},
  pages={035440},
  year={2015},
}

@article{silva2020electronic,
  author={Silva, JEG and Furtado, Job and Santiago, Thiago M and Ramos, Antonio CA and Da Costa, DR},
  journal={Phys. Lett. A},
  volume={384},
  number={25},
  pages={126458},
  year={2020},
}

@article{yecsiltacs2022dirac,
  author={Ye{\c{s}}ilta{\c{s}}, {\"O}ZLEM and Furtado, Job and Silva, J. E. G.},
  journal={Eur. Phys. J. Plus},
  volume={137},
  number={4},
  pages={1--12},
  year={2022},
}

@article{bueno2012landau,
  author={Bueno, MJ and Furtado, C and de M. Carvalho, AM},
  journal={Eur. Phys. J. B},
  volume={85},
  pages={1--5},
  year={2012},
}

@article{villalba2001energy,
  author={Villalba, V{\i}ctor M and Rinc{\'o}n Maggiolo, A},
  journal={Eur. Phys. J. B},
  volume={22},
  pages={31--35},
  year={2001},
}

@inproceedings{vozmediano2008gauge,
  author={Vozmediano, Maria AH and de Juan, Fernando and Cortijo, Alberto},
  booktitle={J. Phys.: Conf. Ser.},
  volume={129},
  number={1},
  pages={012001},
  year={2008},
}

@article{abergel2010properties,
  author={Abergel, DSL and Apalkov, V and Berashevich, J and Ziegler, Klaus and Chakraborty, Tapash},
  journal={Adv. Phys.},
  volume={59},
  number={4},
  pages={261--482},
  year={2010},
}

@article{olpak2012dirac,
  author={Olpak, Mehmet Ali},
  journal={Mod. Phys. Lett. A},
  volume={27},
  number={03},
  pages={1250016},
  year={2012},
}

@article{gallerati2019graphene,
  author={Gallerati, Antonio},
  journal={Eur. Phys. J. Plus},
  volume={134},
  number={5},
  pages={202},
  year={2019},
}

@article{shokri2022rkky,
  author={Shokri, Behzad and Eghbali, Ali and Phirouznia, Arash},
  journal={Phys. Rev. B},
  volume={106},
  number={19},
  pages={195426},
  year={2022},
  
}

@article{birrell1984quantum,
  author={Birrell, Nicholas David and Davies, Paul Charles William},
  year={1984},
  publisher={Cambridge university press}
}

@book{katsnelson2020physics,
  author={Katsnelson, Mikhail I},
  year={2020},
  publisher={Cambridge University Press}
}

@article{silva2024strain,
  author={Silva, JEG and Ye{\c{s}}ilta{\c{s}}, {\"O} and Furtado, J and Filho, AA Ara{\'u}jo},
  journal={Eur. Phys. J. Plus},
  volume={139},
  number={8},
  pages={762},
  year={2024},
}

@book{wald2010general,
  author={Wald, Robert M},
  year={2010},
  publisher={University of Chicago press}
}

@article{hayashi2010curvature,
  author={Hayashi, Masako and Inagaki, Tomohiro},
  journal={Int. J. Mod. Phys. A},
  volume={25},
  number={17},
  pages={3353--3374},
  year={2010},
 
}

@article{batista2018curvature,
  author={Batista Jr, FF and Chaves, Andrey and Da Costa, DR and Farias, GA},
  journal={Physica E},
  volume={99},
  pages={304--309},
  year={2018},
}

@article{kothari2019critical,
  author={Kothari, Mrityunjay and Cha, Moon-Hyun and Lefevre, Victor and Kim, Kyung-Suk},
  journal={Proc. R. Soc. A},
  volume={475},
  number={2221},
  pages={20180671},
  year={2019},
}

@article{yan2013strain,
  journal={Nat. Commun.},
  volume={4},
  number={1},
  pages={2159},
  year={2013},
}

@article{yang2012electronic,
  author={Yang, Mou and Cui, Yan and Wang, Rui-Qiang and Zhao, Hong-Bo},
  journal={J. Appl. Phys.},
  volume={112},
  number={7},
  year={2012},
  publisher={AIP Publishing}
}

@article{Matrix_Method,  
  author  = {Amodio, P. and Mazzia, F.},
  journal = {Applied Numerical Mathematics},
  volume  = {55},
  number  = {1},
  pages   = {70--90},
  year    = {2005},
  publisher = {Elsevier},
  doi     = {10.1016/j.apnum.2005.02.008}
}

@article{Fermions_Charuto,
    author = "Dantas, D. M. and Veras, D. F. S. and Silva, J. E. G. and Almeida, C. A. S.",
    eprint = "1506.07228",
    archivePrefix = "arXiv",
    primaryClass = "hep-th",
    doi = "10.1103/PhysRevD.92.104007",
    journal = "Phys. Rev. D",
    volume = "92",
    number = "10",
    pages = "104007",
    year = "2015"
}

@article{Graviton_Charuto,
    author = "Veras, D. F. S. and Silva, J. E. G. and Cruz, W. T. and Almeida, C. A. S.",
    eprint = "1409.3180",
    archivePrefix = "arXiv",
    primaryClass = "hep-th",
    doi = "10.1103/PhysRevD.91.065031",
    journal = "Phys. Rev. D",
    volume = "91",
    number = "6",
    pages = "065031",
    year = "2015"
}

@article{Graviton_Julio,
    author = "Araujo, J. C. B. and Silva, J. E. G. and Veras, D. F. S. and Almeida, C. A. S.",
    eprint = "1410.3164",
    archivePrefix = "arXiv",
    primaryClass = "hep-th",
    doi = "10.1140/epjc/s10052-015-3350-8",
    journal = "Eur. Phys. J. C",
    volume = "75",
    number = "3",
    pages = "127",
    year = "2015"
}

@article{Gauge_Charuto,
    author = "Costa, F. W. V. and Silva, J. E. G. and Veras, D. F. S. and Almeida, C. A. S.",
    eprint = "1501.00632",
    archivePrefix = "arXiv",
    primaryClass = "hep-th",
    doi = "10.1016/j.physletb.2015.06.042",
    journal = "Phys. Lett. B",
    volume = "747",
    pages = "517--522",
    year = "2015"
}

@article{Veras:2017nke,
    author = "Veras, D. F. S. and Almeida, C. A. S.",
    eprint = "1702.06263",
    archivePrefix = "arXiv",
    primaryClass = "gr-qc",
    doi = "10.1103/PhysRevD.95.104032",
    journal = "Phys. Rev. D",
    volume = "95",
    number = "10",
    pages = "104032",
    year = "2017"
}

\end{document}